\documentclass[preprint,5p,times,twocolumn,sort&compress]{elsarticle}

\usepackage{amssymb}
\usepackage{amsmath}

\journal{Nuclear Instruments and Methods in Physics Research A}

\begin{document}

\begin{frontmatter}



\title{Electron cloud density measurements in accelerator beam-pipe \\ using resonant microwave excitation}


\author[Cornell]{J.P. Sikora\corref{cor1}}
\ead{jps13@cornell.edu}

\author[MELLON]{B.T. Carlson}
\author[GORDON]{D.O. Duggins}
\author[COLUMBIA]{K.C. Hammond}

\author[LBNL]{S. De Santis}

\author[IDAHO]{A.J. Tencate}

\address[Cornell]{CLASSE, Cornell University, Ithaca, New York 14853 United States}
\address[MELLON]{Carnegie Mellon University, Pittsburgh, Pennsylvania 15213 United States}
\address[GORDON]{Gordon College, Wenham, Massachussetts 01984 United States}
\address[COLUMBIA]{Columbia University, New York, New York 10027 United States}
\address[LBNL]{LBNL, Berkeley, California 94720, United States}
\address[IDAHO]{Idaho State University, Pocatello, Idaho 83209 United States}

\cortext[cor1]{Corresponding author. Tel.: +1 6072554882}

\begin{abstract}

An accelerator beam can generate low energy electrons in the beam-pipe, generally called electron cloud, that can produce instabilities in a positively charged beam. One method of measuring the electron cloud density is by coupling microwaves into and out of the beam-pipe and observing the response of the microwaves to the presence of the electron cloud. In the original technique, microwaves are transmitted through a section of beam-pipe and a change in EC density produces a change in the phase of the transmitted signal. This paper describes a variation on this technique in which the beam-pipe is resonantly excited with microwaves and the electron cloud density calculated from the change that it produces in the resonant frequency of the beam-pipe. The resonant technique has the advantage that measurements can be localized to sections of beam-pipe that are a meter or less in length with a greatly improved signal to noise ratio.

\end{abstract}


\end{frontmatter}

\section{Introduction}

   An accelerator beam can generate low energy electrons by ionization of the residual gas in the beam-pipe, lost beam particles or from the photo-electrons produced by synchrotron radiation. These electrons can then produce secondary electrons, generating an electron cloud (EC) that may result in instabilities and emittance growth in a positively charged accelerator beam~\cite{FurmanPivi,ECLOUD12:FURMAN}. Several techniques have been developed to measure the EC density in order to test mitigation techniques and for comparison with the results of numerical simulations of cloud development~\cite{PhaseI,ICFA_News:ZIMMERMANN}.
   
One of these techniques couples microwaves into the beam-pipe and uses the interaction of electrons with the microwaves to measure the EC density. When this technique was first introduced, the microwaves were propagated for some distance through a section of beam-pipe.  A change in EC density within this section produces a change in the phase of the transmitted signal that is proportional to the electron cloud density and the  transmission length~\cite{PAC05:MPPP031,PRL100:094801,PRSTAB13:071002,PRSTAB14:012802}. For a periodic EC density, such as that produced by a short train of bunches in a storage ring, this results in a received signal with phase modulation sidebands above and below the drive frequency that can be observed with a spectrum analyzer. These measurements are referred to as microwave transmission measurements or -- since they generally use the fundamental TE mode propagating in the beam-pipe -- the TE wave technique of electron cloud measurement.

In practice, transitions in the cross-section of the beam-pipe for accelerator hardware such as wigglers, gate valves and the longitudinal slots for vacuum pumps produce reflections of the microwaves. In general, such reflections dominate the beam pipe transmission spectrum just above the cutoff frequency. This results in a series of resonances rather than the usual waveguide transmission function, except for cases with long portions of uniform beam-pipe. The calibration of the transmission measurement is a phase shift per unit length, but reflections make the length of propagation difficult to determine.

An alternative to the microwave transmission method is to make use of the resonant response of the beam-pipe when microwaves are coupled into it. The simplest case would be a region of beam-pipe that contains two perfect reflectors. This is equivalent to a length of waveguide with its ends shorted, where the lowest resonant frequencies are given by Eq.~\ref{eq:fsquared}, with $n$ an integer greater than zero, $f_c$ the cutoff frequency of the beam-pipe, $c$ the speed of light and $L$ the length of the resonant section. 

\begin{equation}
 f^2 = f_c^2 + (nc/2L)^2
\label{eq:fsquared}
\end{equation}

\begin{figure}[htb]
   \centering
   \includegraphics*[width=.8 \columnwidth]{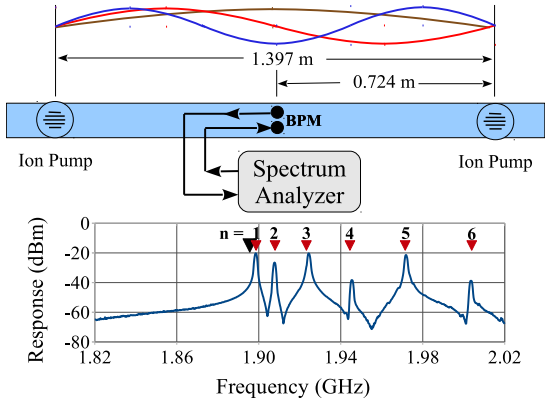}
   \caption{At the location 43E in the C{\scriptsize ESR}TA storage ring, a response measurement shows resonances produced by reflections at two ion pumps. The numbered triangles show the resonant frequencies expected for a shorted section of waveguide of length $L = 1.385$~m. The leftmost dark triangle is the beam-pipe cutoff frequency $f_c$ of 1.8956~GHz. } 
   \label{IPAC11:43E_response}
\end{figure}

Figure~\ref{IPAC11:43E_response} shows an example from the Cornell Electron Storage Ring Test Accelerator (C{\footnotesize ESR}TA)~\cite{ICFA_News:DUGAN} where a response measurement from a section of beam-pipe is consistent with the model of a shorted section of waveguide. At this location, the beam-pipe is of uniform cross section in the region between two vacuum pumps. Above the cutoff frequency of the beam-pipe, longitudinal slots at vacuum pumps generate reflections and a resonant response. The measured resonances are plotted in Fig.~\ref{IPAC11:43E_response} along with the frequencies calculated with  Eq.~\ref{eq:fsquared}, where $L$ is roughly the distance between the ion pumps. For this measurement, the microwaves are coupled in and out of the beam-pipe at the same longitudinal position using electrodes within this section that are normally connected to the beam position monitor (BPM) system. 

When the beam-pipe has a resonant response,  the analysis to determine the EC density within it needs to be based on the response of these resonances to changes in EC density~\cite{IPAC11:SIKORA_TEW,ECLOUD12:SIKORA_TEW,IPAC13:Fermilab}. This paper describes a variation on the microwave transmission method that is based on techniques for plasma density measurement using resonant cavities. This method is referred to as resonant microwave or resonant TE wave measurement of EC density.

\section{Resonant frequency shift due to a plasma}

The natural frequency $\omega_0$ of a resonator will be shifted by an amount $\Delta \omega$ by the presence of a plasma within it. In the absence of a magnetic field, the frequency shift is given by Eq.~\ref{dw:plasmafreq}, where the integrals are taken over the volume of the resonator, $\nu$ is the collision frequency of the plasma and  $\omega_p$  its plasma frequency~\cite{PR106:196}. This approximation is valid for $\Delta \omega / \omega_0$ small, where the electric field in the presence of a plasma is approximately equal to the electric field $E_0$ of the empty cavity. 

\begin{equation} 
\frac{\Delta \omega}{ \omega_0} \; \approx \;
\left( \frac{1}{1 + (\nu / \omega_0)^2 } \right) \left( \frac{1}{2} \right) \frac{\displaystyle \int_{V} \left(  \omega_p^2 / \omega_0^2  \right) E_{0}^2 \,dV  }
       { \displaystyle \int_{V} E_{0}^{2} \,dV }
\label{dw:plasmafreq}
\end{equation}

In accelerators, relatively low EC densities $n_e$ are expected.   
Based on experimental evidence as well as analytical models and simulations, $n_e$ of only about \mbox{ $10^{12}$~$e^{-}/$m$^{3}$} are sufficient to produce beam instabilities~\cite{ECLOUD12:FURMAN,PhaseI,ICFA_News:ZIMMERMANN}. The approximate collision frequency $\nu$ of electrons in a plasma is given in Eq.~\ref{collisionfreq} with temperature $T$ in Kelvin~\cite{LAArzimovich1965:ElemPlasmaPhys}. 

\begin{equation}
\nu \approx      \frac {3.75 \times 10^{-6}  n_e}{T^{3/2}}
\label{collisionfreq}
\end{equation}

If the electrons have an average energy of 1.5~eV, their equivalent temperature is on the order of $10^{4}$~K. An EC density of \mbox{ $10^{12}$~$e^{-}/$m$^{3}$} would have a collision frequency of under 40~Hz. So under these conditions $\nu \ll \omega_0$ and $\nu / \omega_0 \rightarrow 0$. There is also a change in the resonator Q at higher plasma densities, but this change is roughly proportional to $\nu / \omega_0$, so changes in Q will not be treated here. 

The plasma frequency $\omega _{p}$ is related to the plasma (electron cloud) density $n_e$ by $ n_{e}  = \omega _{p}^{2} \varepsilon _0 m_e / e^{2} $ \cite{MAHeald1965:PlasDiagMicroW}, where $\varepsilon _0$ is the vacuum permittivity, $m_e$ is the mass and $e$ the charge of an electron. Combining this with Eq.~\ref{dw:plasmafreq} for a low density plasma ($ \nu / \omega_0 \rightarrow 0$) in the absence of an external magnetic field, the change in resonant frequency for a given EC density $n_e$ is given by Eq.~\ref{dw:Ne}, where the local EC density $n_e$ can be a function of time and of the position within the resonator. This expression and simulations of electron cloud in resonant beam-pipe are in good agreement~\cite{NIM14:SONNAD}.

\begin{equation}
\frac{\Delta \omega}{ \omega_0} \; \approx \;
 \frac{e^{2} }{2 \varepsilon _{0} m_{e} \omega_0^{2} } \frac{ \displaystyle \int_{V} n_{e} E_{0}^{2}\,dV  }
       { \displaystyle \int_{V} E_{0}^{2} \,dV } 
	\label{dw:Ne}
\end{equation}

\section{Calculation of electron cloud density}
\label{Calculation}

Equation~\ref{dw:Ne} is the basis for the resonant microwave technique for measurement of EC density. The frequency shift  $\Delta \omega$ is related to the EC density $n_e$ through a ratio of integrals where the numerator generally requires knowledge of the product of $n_e$ and $E_0^2$ everywhere within the resonant volume at any given time.

A detailed evaluation of Eq.~\ref{dw:Ne} includes determining the distribution of microwaves $E_0^2$ within the resonant section of beam-pipe. With rectangular or elliptical beam-pipe, the transverse distribution of the electric field at the lowest resonant frequencies will be a half wavelength cosine with a maximum in the horizontal center of the beam-pipe. Longitudinally,  unless the series of resonant frequencies is very simple as in the example of Fig.~\ref{IPAC11:43E_response}, the length of the resonant section and the electric field distribution may not be obvious. Some estimates need to be made for the length of the resonant volume in order to determine the possible variation of $n_e$ within that volume. For example, if the resonant section were known to be short, it might be easier to make an approximation that $n_e$ is uniform within that section. Comments on the determination of  $E_0^2$ are  deferred to Section~\ref{BeadPull}. 

The EC density $n_e$ is generally not uniform over the volume. For example, where synchrotron radiation is present, the production rate of initial photo-electrons will vary with the distance from the radiation source. This will result in a corresponding change in the overall EC density as a function of longitudinal position. The EC density can also vary over the transverse dimensions of the beam-pipe.  Calculation or simulation of EC density as a function of position in the beam-pipe is beyond the scope of this paper.  In order to more easily connect $n_e$ with a measured frequency shift, the approximation is made that at any instant in time, $n_e$ is uniform over the resonant volume. With this approximation, the ratio of integrals simplifies to the spatially uniform EC density $n_e$ that can be a function of time. 

The EC density obtained in this way can also be interpreted as the spatially averaged value, weighted by $E_0^2$. If the fundamental mode is excited, this will have maximum $E_0^2$ at the horizontal center of the beam-pipe which usually coincides with the position of the beam. So the measurement would be of practical interest since it measures the average EC in the region of the beam-pipe that includes the beam.

To determine the EC density, the resonant frequency shift $\Delta \omega$ or quantities that are proportional to that shift are measured. From Eq.~\ref{dw:Ne}, if the EC density is spatially uniform and changes from zero to some fixed value $n_e$, the frequency shift produced by that change will be

\begin{equation}
\frac{\Delta \omega}{ \omega_0} \; \approx \;
 \frac{e^{2} }{2 \varepsilon _{0} m_{e} \omega_0^{2} } n_{e}.
	\label{dw:NeConstant}
\end{equation}
 
\subsection{Steady state changes in resonant frequency} 
 
A spectrum analyzer can be used to obtain the beam-pipe response versus frequency near resonance and measure the change in resonant frequency $\Delta \omega$ when an EC density is present. The accuracy of this measurement is determined by the ability to measure the peak response frequency. This will be some fraction of the -3~dB (full) width of the resonance $\Delta \omega_{width} / \omega_0 = 1 / Q$. An estimate for the minimum measurable EC density is obtained by setting the measured  $\Delta \omega / \omega_0$ of Eq.~\ref{dw:NeConstant} equal to the -3~dB width of the resonance. For example, with a $Q$ of 3000 and frequency of 350~MHz, an EC density of \mbox{$10^{12} e^{-}/$~m$^{-3}$} would be easily measurable. This technique measures the time-averaged frequency shift and is best suited for measurement of a step change in EC density that lasts at least several seconds. One practical difficulty is that other effects, such as changes in the temperature of the beam-pipe, will also change its resonant frequency. So experiments need to be designed to separate electron cloud from other effects. An example of a measurement of this type can be found in Ref.~\cite{IPAC12:Alesini}, where clearing electrodes are used to change the EC density.

\subsection{Periodic changes in resonant frequency}
\label{periodic}

Under conditions where the EC density changes with time, some approximation for the time variation of $n_e$ is needed. When the EC density is periodic, the resonant frequency will follow the instantaneous EC density and also be periodic. A method for detecting this periodic frequency shift is to excite the beam-pipe at a fixed frequency that is at or very close to a resonance. 

The equilibrium response of an oscillator driven at a fixed frequency $\omega$ close to a particular resonant frequency $\omega_0$ is given by Eq.~\ref{eq:x(t)} where the resonator quality factor Q  is related to its damping time $\tau$ by $Q = \omega_0 \tau / 2$ and the constant $A$ is proportional to the drive amplitude.  Equations~\ref{eq:An} and~\ref{eq:phi_n} are the equilibrium amplitude and phase response near this resonant frequency (as in Ref~\cite{KRSymon1971:Mechanics}).

 \begin{eqnarray}
 x(t)    & = & A_0  sin(\omega t + \phi_0 )      \label{eq:x(t)} \\ 
 \nonumber \\
where \nonumber \\
A_0     & = &   Q  \frac{A}{[ Q^2(\omega_0^2 - \omega^2)^2 +  \omega^2 \omega_0^2 ]^{1/2} }  \label{eq:An}  \\
\nonumber \\
\phi_0 & = & tan^{-1} \left[ Q \frac{ (\omega_0^2 - \omega^2) }{ \omega \omega_0 } \right] \label{eq:phi_n}
\end{eqnarray}

\begin{figure}[htb]
   \centering
   \includegraphics*[width=.8 \columnwidth]{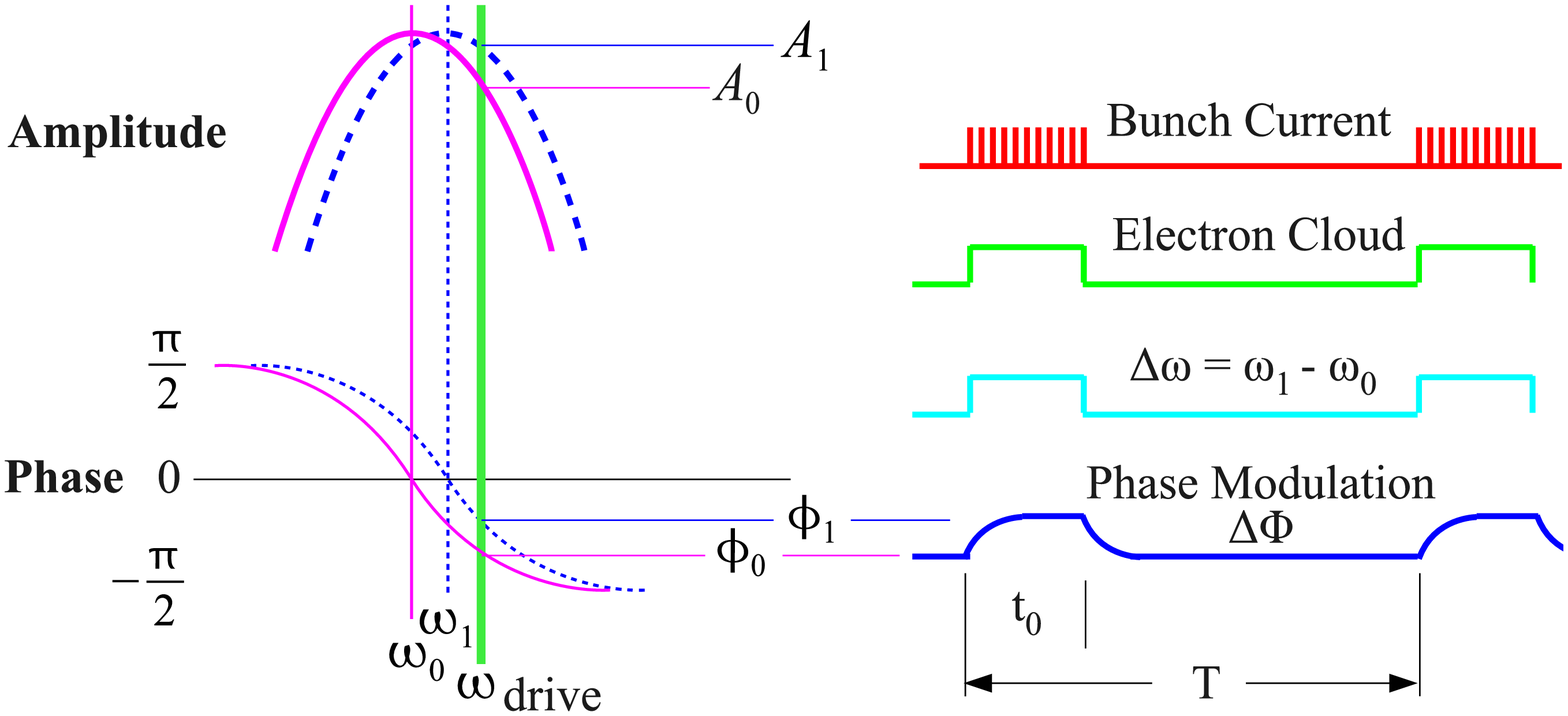}
   \caption{The equilibrium amplitude and phase response of a resonance versus drive frequency are shown at left, corresponding with Eqs.~\ref{eq:An} and \ref{eq:phi_n}. At right is a sketch of a bunch train with duration $t_0$ that is short compared to the revolution period $T$ of the train in a storage ring. This train will produce an electron cloud and the resulting frequency shift $\Delta \omega$ and phase shift $\Delta \Phi$.}
   \label{TEW:ampl_phase}
\end{figure}

A change in EC density shifts the resonant frequency from $\omega_0 \rightarrow \omega_1$. With a fixed drive frequency $\omega_{drive}$, this results in a change in amplitude ($A_0 \rightarrow A_1$)  and phase ($\phi_0 \rightarrow \phi_1$) of the received signal, as shown on the left side of Fig.~\ref{TEW:ampl_phase}.

 The right side of  Fig.~\ref{TEW:ampl_phase} illustrates accelerator conditions that could produce a shift of $\Delta \omega = \omega_1 - \omega_0$. A short train of bunches of length $t_0$ in a storage ring with revolution period $T$ produces a periodic EC density and a corresponding shift in the resonant frequency. Modulation of the resonant frequency will generate amplitude and phase modulation of the drive frequency that are observable in its spectrum as modulation sidebands. 


As long as the excitation is very close to resonance, the effects of amplitude modulation only become relevant at much higher EC densities, so the spectra calculated here are based only on the phase modulation of the signal.  Frequency modulation can also be an important component of the spectrum, but additional data is needed to confirm this part of the analysis~\cite{IPAC12:DESANTIS,IBIC13:DESANTIS}. When the EC density is changing, we can identify two different regimes, depending on whether or not the EC density changes are on a time scale that is long compared with the resonance damping time.

\subsubsection{Slow changes in EC density}
\label{slow}

If changes in resonant frequency are slow compared with the resonance damping time, the phase difference between the drive and the resonant microwaves  will remain in equilibrium. Evaluating Eq.~\ref{eq:phi_n} for a small change in resonant frequency, the change in phase with a fixed drive frequency is $\Delta \phi \approx 2 Q \Delta \omega / \omega$. Combining this with the frequency shift of Eq.~\ref{dw:NeConstant} gives the equilibrium phase shift produced by a change in EC density $n_e$.

\begin{equation}
 \Delta \phi  \; \approx \;   \frac{ Q}{\omega_0^2} \frac{e^2}{ \varepsilon_0 m_e} n_e.
\label{eq:deltaphi}
\end{equation}

When the changes in $n_e$ are slow compared with the resonance damping time, they can be measured directly through the changes in the phase of the received signal. 

\subsubsection{Rapid changes in EC density}
\label{rapid}

If there is a step change in resonant frequency, the phase difference between the drive and the resonant microwaves will require several damping times to come into equilibrium. More generally, if the changes in resonant frequency are \textit{not} slow compared with the resonance damping time $\tau$, the transient phase response is obtained by convolving the equilibrium phase of Eq.~\ref{eq:deltaphi} with the $e^{-t/\tau}$ impulse response of the resonance~\cite{IBIC13:SIKORA}.

For convenience, the EC density as a function of time $n_e(t)$ can be written as a the product of the peak EC density  $ n_{peak}$ during that time interval and a function $\rho(t)$ that has a peak value of one. The normalized function $\rho(t)$ can then be scaled when calculating the peak EC density, for example as a function of beam current, with the assumption that the function will have the same shape over some range of current. When calculating spectra in Section~\ref{calculating}, $\rho(t)$ is convolved with the resonance damping time and Fourier components are calculated that can also be scaled by the peak EC density.

Define the quantity $\Delta \phi_{peak}$ proportional to the peak EC density.  

\begin{equation}
\Delta \phi_{peak} \equiv  \frac{Q e^2} { \omega_0^2 \varepsilon_0 m_e} n_{peak}
\label{eq:phi_peak}
\end{equation}

Then the phase $\Delta \Phi$, that is proportional to the convolution of $\rho(t)$ with the impulse response can be written as

\begin{eqnarray}
\Delta \Phi (t)  =&  \frac{ Q}{\omega_0^2} \frac{e^2}{ \varepsilon_0 m_e}  \int_{-\infty}^t n_e(\xi) e^{(t - \xi)/\tau}  d \xi \ \ \ \ &  \label{eq:Phi_conv}    \\
  =&   \Delta \phi_{peak}  \int_{-\infty}^t \rho(\xi) e^{(t - \xi)/\tau}  d \xi . \ \ \ \ &  \nonumber 
 \end{eqnarray}

\subsubsection{Calculating EC density from spectra}
\label{calculating}

Changes in EC density $n_e$ can be measured by detecting the phase shifts in received signals. One technique for doing this is to detect the phase using a mixer, where the drive signal is multiplied by the received signal and the phase difference observed in the time domain with an oscilloscope as described in Ref.~\cite{IBIC13:phase_det}. However, most of our measurements are made in the frequency domain, observing the spectra produced by periodic changes in phase. 

If the magnitude of the phase modulation is small, it is easy to quantify the magnitude of the phase modulation sidebands that are generated.  A drive signal at frequency $\omega$ with a small cosine phase modulation of magnitude $M \ll 1$ and modulation frequency $\omega_{T}$ , can be written as in Eq.~\ref{eq:sidebands}. Using trigonometric identities, the result is a signal at the drive frequency $\omega$ and sidebands at frequencies $\omega_{T}$ above and below the drive frequency, both sidebands having an amplitude $M/2$ with respect to the drive. 

\begin{eqnarray} 
g(t) &= & \sin[ \omega t + M \cos(\omega_{T} t) ] \nonumber \\
      & \approx & sin(\omega t)                              \nonumber \\
      &    & + \frac{M}{2} \left[ \cos( \omega + \omega_{T})t + \cos( \omega - \omega_{T})t \right]
      \label{eq:sidebands}
\end{eqnarray} 

Any periodic phase modulation, such as that described by Eq.~\ref{eq:Phi_conv}, can be expressed as a Fourier series $\Delta \Phi = 
\Delta \phi_{peak}  \sum_{m=1}^{+\infty}C_m \cos(m \omega_T t  )$  and each component of that series will produce a pair of sidebands. The phase of each component has been dropped since there is only one component at each sideband frequency and only the magnitude of the sideband is needed to calculate the measured spectrum.  

 \begin{eqnarray} 
 \label{eq:sidebandsCm}
\lefteqn{ g(t) = }   \\   
&& \sin \left[ \omega t + \Delta \Phi (t) \right]   \nonumber \\
&=& \sin \left[ \omega t + \Delta \phi_{peak}  \int_{-\infty}^t \rho(\xi) e^{(t - \xi)/\tau}  d \xi     \right]   \nonumber \\
      &= &  \sin \left[ \omega t + \Delta \phi_{peak} \sum_{m=1}^{+\infty}C_m \cos(m \omega_T t) \right]  \nonumber \\
      & \approx & sin(\omega t) \nonumber \\
      && + \Delta \phi_{peak} \sum_{m=1}^{+\infty} \frac{C_m}{2} \left[ \cos(\omega + m \omega_T)t +\cos(\omega - m \omega_T)t  \right]  \nonumber      
\end{eqnarray}

If the normalized shape of the modulation $\rho(t)$ is known, the Fourier coefficients $C_m$ of Eq.~\ref{eq:sidebandsCm} can be calculated. Define the sideband ratio $S_m$ to be the ratio of the sideband amplitude to the drive amplitude.  Using Eq.~\ref{eq:phi_peak}, this is

\begin{equation}
S_m \approx  \frac{1}{2}\Delta  \phi_{peak} C_m   \approx  \frac{ Q}{\omega_0^2} \frac{e^2}{ \varepsilon_0 m_e} \frac{C_m}{2} n_{peak}.
\label{eq:Sm}
\end{equation}

Rearranging this equation, the measured sideband ratio $S_m$ and the Fourier components $C_m$ can be used to calculate the peak EC density $n_{peak}$.

\begin{equation}
n_{peak} \approx  S_m \frac{2  \omega_0^2 }{ Q \cdot C_m} \frac{ \varepsilon_0 m_e }{e^{2}}
\label{eq:ne}
\end{equation}


 For example, if the EC density has a fixed amplitude $n_{peak}$ for a time interval $0 \le t \le t_0$ that is short compared with the period $T$ and is zero otherwise, the convolved phase shift $\Delta \Phi$ is given by Eq.~\ref{eq:rectconv} and illustrated in the sketch of Fig.~\ref{TEW:ampl_phase}.


\begin{eqnarray}
\Delta \Phi (t)   = & \Delta \phi_{peak}  \int_{-\infty}^t \rho(\xi) e^{(t - \xi)/\tau}  d \xi \ \ \ \    \label{eq:rectconv}    \\
                                                                                                                                                                   \nonumber \\
= & \begin{cases}
   \Delta \phi_{peak} (1 - e^{-t/\tau})                                        &            (0 \leq t \leq t_0)   \nonumber  \\
 \Delta \phi_{peak}  (1 -  e^{-t_0/\tau})   e^{-(t-t_0)/\tau}       &            ( t  \geq t_0 )
\end{cases}   \\                   
                                                                                                                                                                   \nonumber \\                                                                                                            
=     &  \Delta \phi_{peak}  \sum_{m=1}^{+\infty}C_m \cos(m \omega_T t)   &   \nonumber 
\end{eqnarray}

The Fourier components of this phase modulation are given by Eq.~\ref{eq:c_m} where the modulation frequency for a single train in a storage ring is $\omega_T = 2 \pi / T$ with $T$ the revolution period of the beam and $\tau = 2 Q/\omega_0$ is the damping time of the beam-pipe resonance.



\begin{eqnarray}
C_m &  = &  \sin  \left(  \frac{m \omega_T t_0}{2} \right) \frac{2}{\pi} \left[ \left(  \frac{1}{m} - \frac{m (\tau \omega_T)^2}{ 1 + m^2( \tau \omega_T)^2 } \right)^2  \right. \nonumber \\
     & & \left.  +\left(  \frac{ (\tau  \omega_T)}{ 1 + m^2 (\tau \omega_T)^2} \right)^2 \right]^{\frac{1}{2} }     
\label{eq:c_m}
\end{eqnarray}

\subsection{Sensitivity and errors}
\label{sensitivity}

The sensitivity of this technique can be understood through Eqs.~\ref{eq:An} and~\ref{eq:Sm}. From Eq.~\ref{eq:An}, the amplitude of the drive signal response on resonance will be proportional to $Q$. From Eq.~\ref{eq:Sm}, for a given change in EC density $n_{peak}$,  the sideband to drive amplitude ratio $S_m$ will also be roughly proportional to $Q$. So for modest values of $Q$, the absolute amplitude of the sidebands scales as $Q^2$. 

With resonances having $Q$s of several thousand, the typical received signal at the drive frequency can be 0~dBm or higher. So the minimum measurable $n_e$, which corresponds to the minimum measurable sideband to drive ratio $S_m$, is generally determined by the dynamic range of the spectrum analyzer rather than its noise floor. 

Sensitivity would be greatest when the value of $C_m$ is maximized. For example, this will occur for the first sideband if the ratio of the train length $t_0$ to the revolution period $T$ were equal to 1/2, since by Eq.~\ref{eq:c_m}, $C_1$ is proportional to $\sin(\pi t_0/T)$ . 

For example at C{\footnotesize ESR}TA, typical values would be: $Q = 3000$, {\mbox {$\omega_0 = 2 \pi \cdot$(2~GHz)}}, and $C_1$ = 0.06 for a 126~ns long 10-bunch train with a 2562~ns revolution period. We have used an Agilent MXA N9020 with a dynamic range of 78~dB. So if  -78~dBc sidebands are measured under these conditions, the approximate minimum measurable value of $n_e$ would be $7 \times 10^{10}$~$e^{-}/$m$^{3}$.

Measurement errors are dominated by measurement of the $Q$ of the resonance and the value of Fourier components $C_m$.  We have measured the $Q$ using the -3dB amplitude points of the response, but the frequencies of these points are not quite symmetric above and below the resonance peak, probably due to the influence of nearby resonances giving errors of about $\pm$5~\%. Measurement of the sideband amplitude ratio has an estimated error of about $\pm$2\%. Errors in the value of $C_m$ are determined by our approximation for the time evolution of $n_e$ and on the $Q$. For our measurements in next section, our total measurement error is about $\pm$12\%. This error could certainly be improved, especially through a better measurement of the Q. 

Another possible source of error is the approximation used in Eq.~\ref{eq:sidebands} that results in a small phase modulation M giving a sideband to drive ratio $S_m = M/2$. For larger modulations, a more correct value would be the ratio of Bessel functions $S_m = J_1(M)/J_0(M)$. For a value of $S_m = 0.05$,  the $M/2$ approximation would have an error of about 0.1~\%; a value of $S_m = 0.5$  would have an error of about 15~\%. In addition, with sideband amplitudes that are this large, the linear approximation of the phase shift used in Eqs.~\ref{eq:deltaphi} through \ref{eq:c_m} would no longer be valid. So greater care would have to be taken in the analysis with $S_m$ values greater than about 0.05.

The original microwave transmission method and this newer resonant method have important differences in sensitivity.  As can be seen from Eq.~\ref{eq:An}, if the same coupling into and out of the beam-pipe is used with the same drive amplitude, the signal level of the resonant response will be $Q$ times that of the transmission method. This gives a distinct advantage in signal to noise for the resonant method. However, if the interest is in measuring the time development of the electron cloud, a high Q becomes a disadvantage since the phase response can be dominated by the resonance damping time rather than the electron cloud growth and decay time. This is the result of the convolution of Eq.~\ref{eq:Phi_conv}, that tends to smooth rapid changes in phase as shown in the example of  Eq.~\ref{eq:rectconv}. As a result, the resonant method will be less sensitive to details of the time evolution of the electron cloud.

In practice, one is not often given the choice between making a transmission or a resonant measurement; this is determined by the response of the existing beam-pipe. If resonances are observed, the transmission model is not appropriate for calculating the EC density.

\section{EC density measurements}

\begin{figure}[htb]
   \centering
   \includegraphics*[width=.8 \columnwidth]{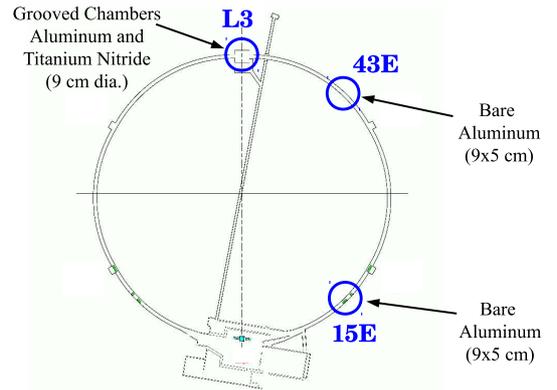}
   \caption{This sketch of the 768~m circumference  C{\scriptsize ESR}TA storage ring shows three locations that are discussed in the text. Most of the beam-pipe is a \mbox{$9 \times 5$~cm} aluminum extrusion. Special chambers were constructed in L3 with 9~cm diameter extrusions. }
   \label{TEW:cesr}
\end{figure}

Cornell's C{\footnotesize ESR}TA storage ring, shown in Fig.~\ref{TEW:cesr}, supports positron and electron beams with bunch charges of up to $1.6 \times 10^{11}$~particles/bunch with energies ranging from 2.1~GeV to 5.3~GeV.  In this storage ring, the source of electron cloud is almost exclusively from synchrotron radiation. This section gives some examples of resonant microwave measurements that were made at Cornell.

There are some practical considerations when making measurements in the presence of a beam signal.  With a bunch charge of  25~nC, the peak voltage induced by the beam on the BPM button electrodes is about 70~V. Precautions must be taken both to protect the instruments and to prevent the beam-induced signal from interfering with the desired microwave measurements. A drive power of 5~W is used to improve signal to noise without risk of damage to the buttons. Bandpass filtering is used on the received signal to limit the peak voltage seen by the spectrum analyzer.  The beam generates harmonics at multiples of the beam revolution frequency. Drive frequencies are chosen that are as close as possible to the beam-pipe resonances, but with both the drive and sidebands of the microwaves falling between the revolution harmonics produced by the beam. 

Three measurements are described in the following sections. In the first, only one resonant frequency is excited and multiple sidebands are recorded  above and below the drive frequency. This data is compared with the spectral envelope that would be expected from phase modulation sidebands as outlined in Section~\ref{calculating}. At the same location, a second measurement uses only the first sidebands, but the measurement is made at five different beam-pipe resonances. The third measurement is at a different location in the storage ring where the electric field distribution had been measured prior to the installation of this section of beam-pipe. Localized resonances allow the comparison of EC density in a bare aluminum chamber with a chamber having a coating of titanium nitride on its vacuum surface.  

In making these measurements, the simplest approximation is used -- that the EC density has a fixed value when the accelerator beam is present at the measurement location and is zero otherwise. The step change in EC density is rapid and must be convolved with the resonance damping time to obtain the correct phase modulation spectra as in Eqs.~\ref{eq:rectconv} and \ref{eq:c_m}. 

\subsection{Hardware setup}

\begin{figure}[htb]
   \centering
   \includegraphics*[width=.8 \columnwidth]{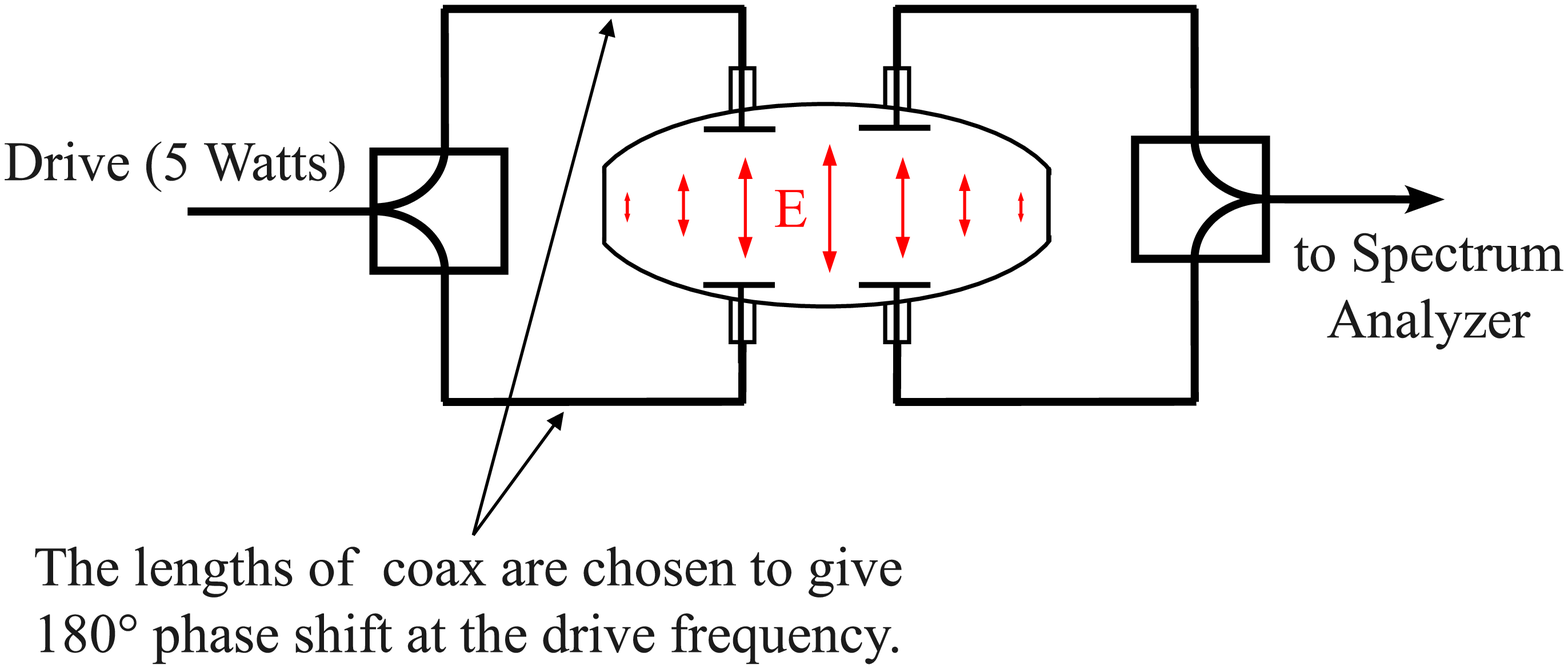}
   \caption{Microwaves excite the beam-pipe in the fundamental waveguide mode using beam position monitor button electrodes. }
   \label{TEW:drive_pickup}
\end{figure}

Data is taken by coupling microwaves into the beam-pipe, typically using pairs of 18~mm diameter BPM buttons that are driven differentially to couple into the desired mode. Differential drive can be obtained using a 180$^\circ$ splitter/combiner and two equal lengths of cable  connected to the buttons. An alternative uses a 0$^\circ$ splitter and two cables of different lengths such that the two buttons are driven with opposite phases at the drive frequency.  With elliptical beam-pipe, a TE mode is driven with a vertically oriented electric field as shown in Fig.~\ref{TEW:drive_pickup}. In round beam-pipe, both horizontal and vertical orientations of the field are possible since they have nearly the same cutoff frequency. All of the data that we have taken so far has used the fundamental mode, with the larger transverse dimension of the beam-pipe being approximately one half wavelength of the excitation frequency.

\begin{figure}[htb]
   \centering
   \includegraphics*[width=.8 \columnwidth]{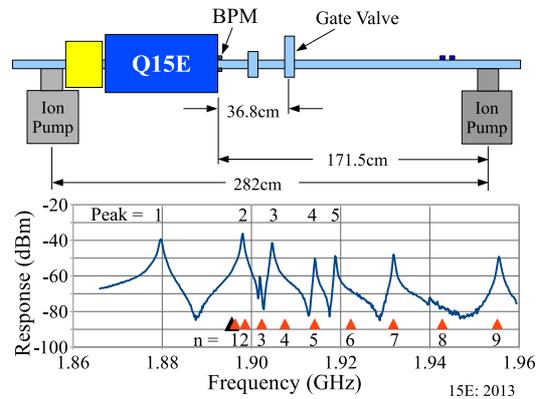}
   \caption{ Five resonances at the 15E location of C{\scriptsize ESR}TA are used to measure the EC density in that section of beam-pipe. The numbered triangles show the resonant frequencies expected for a shorted section of waveguide of length $L = 2.82$~m. The leftmost dark triangle is the beam-pipe cutoff frequency $f_c$ of 1.8956~GHz. There is poor agreement between the calculated and measured values, in contrast to the good agreement of Fig.~\ref{IPAC11:43E_response}  } 
   \label{TEW:15E_response}
\end{figure}

Figure~\ref{TEW:15E_response} shows the resonant response at the location 15E in C{\footnotesize ESR}TA with a bare aluminum chamber using a drive amplitude of -10~dBm. An estimate of the coupling to a particular mode can be made by using the measured insertion loss. For example, the response of Fig.~\ref{TEW:15E_response} shows the first peak at -40~dBm with a drive of -10~dBm. This insertion loss of -30~dBm gives a scattering parameter $S_{21}$ of about 0.032. It can be shown that $S_{21}$ is related to the input and output coupling coefficients $\beta_{in}$ and $\beta_{out}$ by~\cite{ELGinzton1957:MicroMeas}

\begin{equation}
S_{21} = \frac {2 \sqrt{\beta_{in}  \beta_{out}}}{1 + \beta_{in} +  \beta_{out}}.
\end{equation}
 
By symmetry of the button geometry and the resonant mode, the input and output coupling coefficients will be approximately equal. Rearranging terms, Eq.~\ref{beta} gives the value of the coupling, which in this example would be about 0.016. So this arrangement of hardware results in weak coupling to the resonances.

\begin{equation}
\beta_{in}  =      \beta_{out}  =  \frac {S_{21} }{2(1 - S_{21}) }
\label{beta}
\end{equation}

The response at 15E in Fig.~\ref{TEW:15E_response} is similar to that of 43E in Fig.~\ref{IPAC11:43E_response}, in that resonances are established between two ion pumps where longitudinal slots produce reflections. But at the 15E location, the measured resonant frequencies do not agree with the values calculated using Eq.~\ref{eq:fsquared}.  The calculation uses the same cutoff frequency $f_c$ of  1.8956~GHz as used in Fig.~\ref{IPAC11:43E_response}, since the the beam-pipe is from the same aluminum extrusion. The length $L$ was taken to be the measured 2.82~m distance between the ion pumps. Notably, the first peak is below the nominal cutoff frequency of beam-pipe and too far below the other peak frequencies to make a fit with Eq.~\ref{eq:fsquared} possible. It is likely that the vacuum gate valve introduces a locally lower cutoff frequency within its volume and contributes to the displacement of the resonant frequencies, but this has not been confirmed. Some remarks on determining the $E_0^2$ field distribution of beam-pipe are given in Section~\ref{BeadPull}.

\begin{figure}[htb]
   \centering
   \includegraphics*[width=.8 \columnwidth]{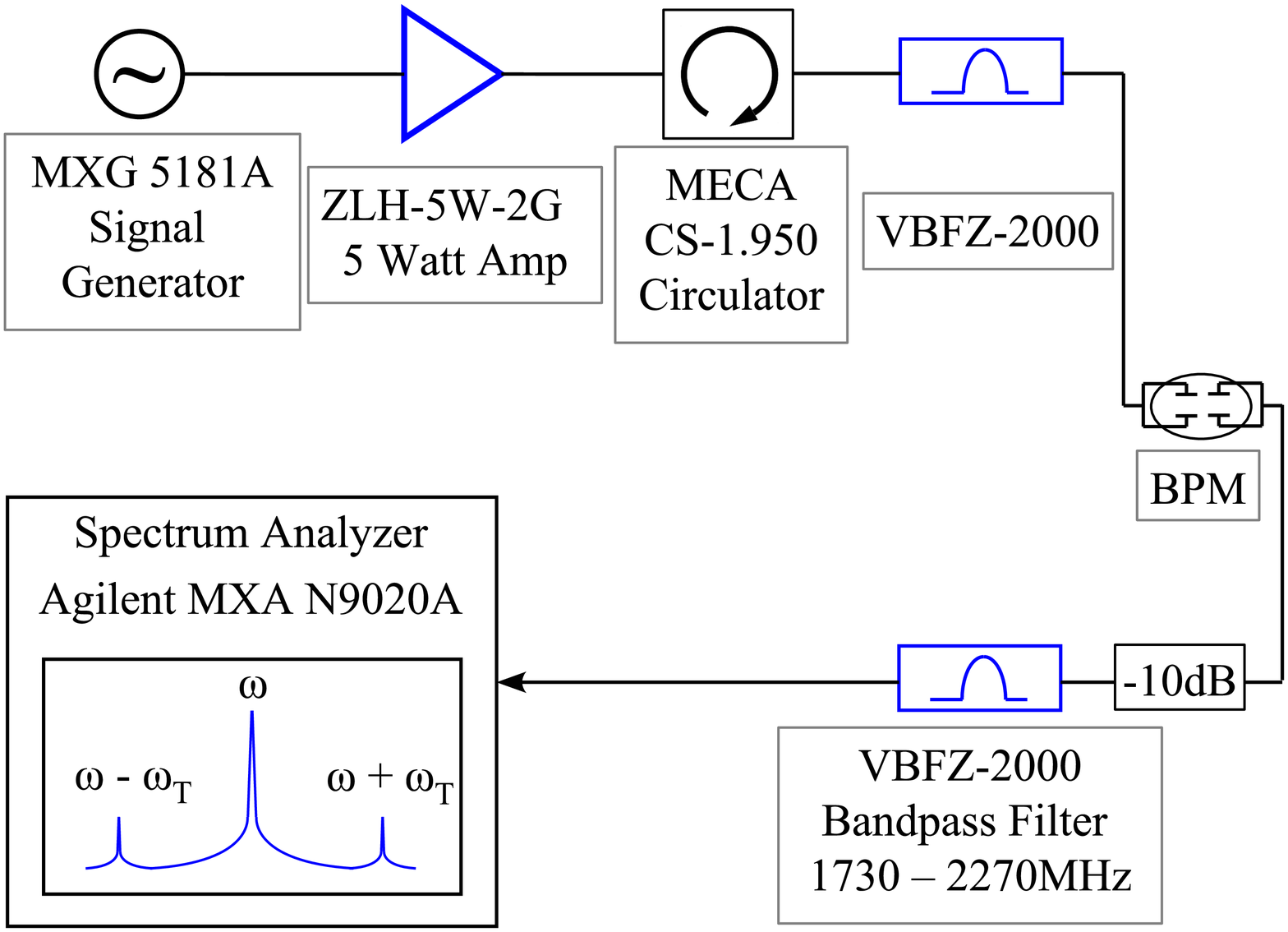}
   \caption{The hardware for TE wave measurements includes a signal generator to excite the beam-pipe and a spectrum analyzer to measure the response.  }
   \label{TEW:schematic}
\end{figure}

Figure~\ref{TEW:schematic} shows the hardware configuration used for measurements made with beam.  A drive signal is amplified to the level of roughly 5~W, and passes through a circulator and bandpass filter before being routed to the beam-pipe with low loss cable. The response of the beam-pipe to this drive signal is detected with a second pair of BPM buttons, routed through a 10~dB attenuator and bandpass filter before being recorded with a spectrum analyzer. The bandpass filters and attenuator serve to protect the equipment from the large, wideband beam-induced signal. Resonance peaks are found by scanning the drive frequency within about 100~MHz of the beam-pipe cutoff frequency and observing the response with the spectrum analyzer -- typically without beam in the storage ring. Drive frequencies are chosen that are on or very close to the observed resonances.

\subsection{Comparing a measured spectrum with calculation }

At C{\footnotesize ESR}TA,  damping times of the beam-pipe resonances range from 400 to 1000~ns, while changes in EC density are generally on a shorter time scale. So convolution of the phase shift with the impulse response of the resonance is needed in order to accurately calculate the phase modulation, as outlined in Section~\ref{rapid}, and the resulting spectra of Section~\ref{calculating}.  

\begin{figure}[htb]
   \centering
   \includegraphics*[width=.8 \columnwidth]{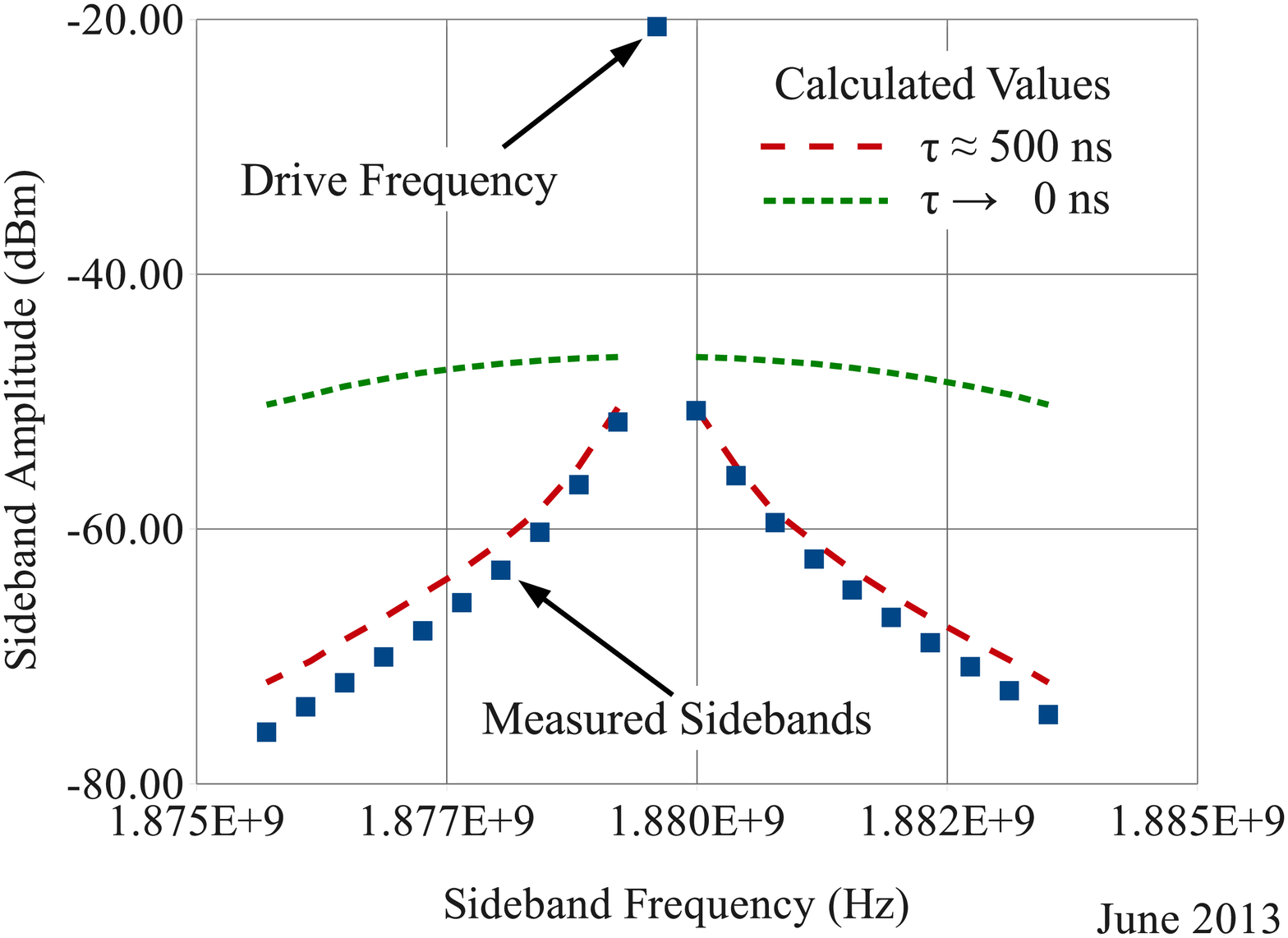}
   \caption{With a 10-bunch train of positrons, measured sidebands are shown as squares with a drive frequency close to the first resonance of Fig.~\ref{TEW:15E_response}. Calculated sideband envelopes from phase modulation are also shown, including convolution with resonance damping times of 500~ns and  0~ns.  }
   \label{TEW:15E_Multiple}
\end{figure}

Figure~\ref{TEW:15E_Multiple} shows the drive and sideband amplitudes measured at 15E with a 10-bunch train of positrons having 14~ns spacing (train length 126~ns) and a beam energy of 5.3~GeV. The bunch charge is $10^{11}$~positrons/bunch (60~mA total current) and the drive frequency is close to the first resonance of Fig.~\ref{TEW:15E_response}, just below 1.88~GHz. 

The 10-bunch train is short compared with the 2562~ns revolution period of the storage ring. Using the approximations of Eq.~\ref{eq:rectconv}, where the EC density has a fixed value for the 126~ns length of the train and is zero otherwise, the envelope of the sidebands is calculated using Eq.~\ref{eq:c_m} and shown in Fig.~\ref{TEW:15E_Multiple}. Two calculated envelopes are shown, the first uses the measured damping time of 500~ns. The second envelope   is for a resonance damping time that approaches zero, where the phase shift follows the step changes in EC density instantaneously.  Using Eq.~\ref{eq:ne},  an EC density of  $4 \times 10^{13}$~$e^{-}/$m$^{3}$ gives a match at the first ($m = 1$) sidebands. For larger values of $m$, the sideband envelope that includes convolution with 500~ns remains within about 2~dB of the measured values up to $m = 10$. In contrast, the spectral envelope that corresponds to $\tau \rightarrow 0$ is not consistent with the measured data. This illustrates the importance of convolution with the resonance damping time when calculating spectra. 
   
 \subsection{EC density versus beam current}
  
Data has also been taken at 15E where the drive frequency is set to each of the five peak frequencies shown in Fig.~\ref{TEW:15E_response} and only the first ($m = 1$) upper and lower sidebands are measured along with the amplitude of the drive (carrier) frequency. The spectrum analyzer returns measurements in decibels. The ratio of the sideband to drive amplitudes $S_m$  is obtained by  $S_m = 10^{dBc/20}$, where $dBc$ is the difference between the drive and sideband amplitudes in decibels. The EC density is calculated from the measured ratio $S_m$ and the measured Q of each resonance, assuming that $n_e$ is uniform over the volume.  Equation~\ref{eq:ne} is used with the sideband index $m$ is always equal to one; $C_m$ is the Fourier coefficient of the convolved phase modulation using the approximations leading to Eqs.~\ref{eq:rectconv} and \ref{eq:c_m}.

\begin{figure}[htb]
   \centering
   \includegraphics*[width=.8 \columnwidth]{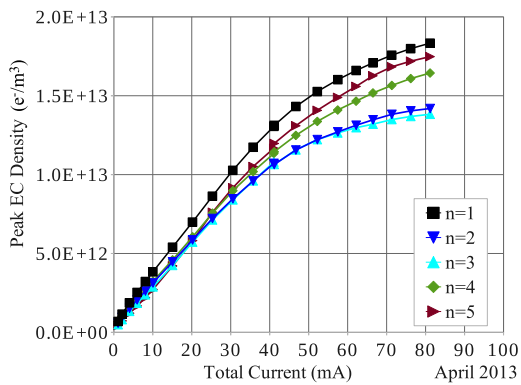}
   \caption{ The EC density $n_e$ is calculated from the ratio $S_1$ for each of the five peak frequencies of Fig.~\ref{TEW:15E_response} and plotted as a function of total beam current in a 10-bunch train of positrons. Relative errors between measurements are about $\pm$12\%, mostly due to errors in the measurement of Q. }
   \label{TEW:15E_Aluminum}
\end{figure}

Data was taken with a 10-bunch train of 5.3~GeV positrons with 14~ns spacing at beam currents up to 80~mA total ($1.28 \times 10^{11}$~positrons/bunch). The EC density for each resonance is calculated to produce the plot in Fig.~\ref{TEW:15E_Aluminum}. At the highest currents of this measurement, the EC density appears to saturate. The maximum $S_m$ of this measurement is roughly 0.040, which is within the range of validity of linear approximations discussed in Section~\ref{sensitivity}.

If the EC density were uniform over the length of this section of beam-pipe, the data obtained from these five resonances should coincide. But at the highest current, the measured EC densities vary by about 30~\%. The differences in these measurements are too large to be explained by errors in the measurement of the $Q$ of each resonance or of the sideband ratios $S_1$. It suggests that the EC density is not uniform and that the distribution of the standing waves is not symmetric. This points out the importance of knowing the distribution of standing waves within a resonant section.

\section{Field distribution of a resonance}
\label{BeadPull}

The EC density will generally not be uniform over the volume, especially when sections of the vacuum surface are coated with materials intended to minimize the production of electron cloud. As suggested at the beginning of Section~\ref{Calculation}, when the EC density varies with position in the chamber it important to know the field distribution of the resonance since from Eq.~\ref{dw:Ne} this determines the contribution of each part of the chamber to the measurement. It is difficult to infer the field distribution of a resonance except in the simplest cases, such as when the series of resonances follows the $f^2 = f_c^2 + (nc/2L)^2$ of a shorted section of waveguide. If this is not the case, modeling of the detailed geometry of the section of beam-pipe would be needed, or some method of measuring the effect of the more complex geometry on the resonant fields. 

\subsection{Bead pull measurement of the L3 beam-pipe}

We measured the resonant field distribution of an assembly of four chambers before they were installed in the L3 section of the storage ring~\cite{IBIC12:L3_beadpull}. The 60~cm long chambers have two different cross sections: one round and the other round with grooves on the upper and lower vacuum surfaces. One of each geometry has a vacuum surface of bare aluminum and the other pair has coatings of TiN. The final assembly has the grooved and smooth chambers alternating.

\begin{figure}[htb]
   \centering
   \includegraphics*[width=.8 \columnwidth]{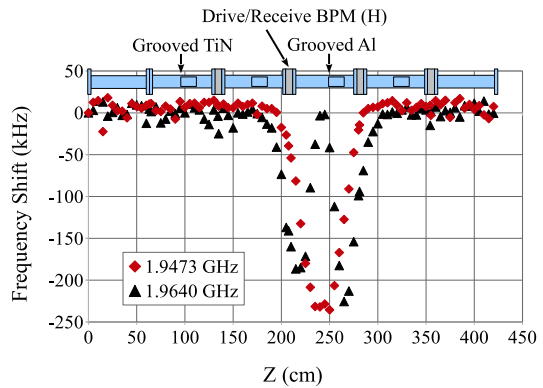}
   \caption{The result of a dielectric bead pull measurement on the four chamber assembly shows two resonances with fields that are primarily within the grooved aluminum chamber. } 
   \label{TEW:beadpull_aluminum}
\end{figure}

To determine the field distribution in the four chamber assembly, a 0.3~cm$^3$ nylon bead was guided along the axis of the beam-pipe and changes in the resonant frequency recorded. The frequency shift is proportional to the square of the electric field at the location of the bead~\cite{BeadPull:Slater,IEEE:Carter}. Figure~\ref{TEW:beadpull_aluminum} shows the result of one of the bead pull measurements, where BPM buttons were used to couple microwaves into and out of the beam-pipe at the same longitudinal position. The measurements show that the resonant field is mostly confined to the 60~cm long aluminum grooved chamber at two frequencies corresponding to $n = 1, 2$. Other measurements gave a similar result for the 60~cm TiN coated grooved chamber at slightly different frequencies and using a different BPM coupler location.  The confinement of the resonant fields is due to the slightly lower cutoff frequency of the grooved chambers compared with their smooth wall neighbors.  The ability to excite these two sections of the chamber independently allows the direct comparison of EC density in the two grooved chambers and the mitigating effects of the TiN coating versus a bare aluminum vacuum surface.

\subsection{EC density measurements in the L3 beam-pipe }
  
\begin{figure}[htb]
   \centering
   \includegraphics*[width=.8 \columnwidth]{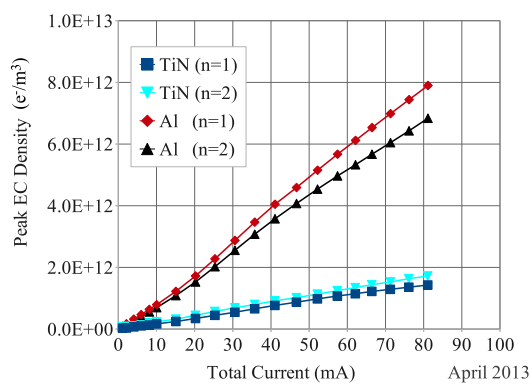}
   \caption{ The EC density versus total beam current was measured in the bare aluminum and TiN coated chambers, showing the mitigating effect of the coating. Two resonances were used in each chamber and relative errors are about $\pm$12\% mostly due to errors in the measurement of Q. }
   \label{TEW:TiN_vs_Al}
\end{figure}

Based on the bench measurements of the four chamber assembly,  BPM couplers and resonant frequencies were selected to make measurements of EC density in the bare aluminum and the TiN coated chambers with beam. Figure~\ref{TEW:TiN_vs_Al} compares the EC density versus total beam current for the these configurations. These two sets of measurements were made at the same time, with high bandwidth coaxial relays used to switch the drive and receive locations between the aluminum and TiN chambers at each beam current. The beam is a 10-bunch train of 5.3~GeV positrons with 14~ns spacing and total beam current ranging from near zero to 80~mA ($1.28 \times 10^{11}$~positrons/bunch). Only the lowest sidebands were recorded and used to calculate the EC density from Eq.~\ref{eq:ne}.

The plots in Fig.~\ref{TEW:TiN_vs_Al} show that a much lower EC density is measured in the TiN coated chamber, even though it is closer to the source of synchrotron light than the bare aluminum chamber. The measurements at the two frequencies in the bare aluminum chamber are within about 10\% of each other, as are those in the TiN chamber. This gives increased confidence that the mitigating effect of the TiN coating can be quantified once the data has been normalized for synchrotron radiation flux. This would not have been possible without the understanding of the distribution of microwaves in these chambers that was obtained through the bead pull measurements.

\section{Summary}

The EC density in accelerator beam-pipe can be measured by resonantly exciting the beam-pipe with microwaves. The sensitivity of the resonant technique is increased by the $Q$ of the resonance, making it much more sensitive than the original microwave transmission technique. However, a high $Q$ also results in less sensitivity to dynamic changes in EC density, since these changes will be convolved with the impulse response of the resonance.    

Calculations of EC density are based on the approximations that the plasma collision frequency $\nu$ is low, magnetic fields can be neglected and that the sidebands are generated by phase modulation of a fixed drive frequency.  When changes in EC density take place on a similar or shorter time scale than the damping time of the resonance, convolution with the resonance impulse response is needed to calculate the phase modulation sidebands. The calculated envelope of the sidebands is in fair agreement with measurements made at C{\footnotesize ESR}TA of the first ten sidebands.  

For convenience,  the calculations and measurements presented here have used the approximation that the EC density is uniform over the resonant volume, but can vary with time. Another approximation is that the EC density has a fixed value when the beam is present in the chamber and zero otherwise.  These approximations make it straightforward to calculate results analytically.

A careful analysis of Eq.~\ref{dw:Ne} would require a more detailed estimate of both the time and spatial distribution of the EC density $n_e$ and of the spatial distribution of the resonant microwaves $E_0$. Bead pull measurements are useful in determining the distribution of the resonant field within the beam-pipe. Measurements have shown resonances confined to sections of beam-pipe less than a meter long. A combination of these techniques has been used to show the mitigating effect of a TiN coating on an aluminum chamber.

\section{Acknowledgements}
This work is supported by the US National Science Foundation \mbox{PHY-0734867} and \mbox{PHY-1002467} and as well as the US Department of Energy \mbox{DE-FC02-08ER41538} and  \mbox{DE-SC0006505.} We would like to thank Yulin Li and the members of the CESR vacuum group for giving us the opportunity to perform bead pull measurements on the L3 vacuum chamber assembly. We are also grateful for the support of the Research Experience for Undergraduates program of the National Science Foundation PHY-0849885 and PHY-1156553.







\begin{thebibliography}{99} 


\bibitem{FurmanPivi}
M.A. Furman and M.T.F. Pivi, ``Probabilistic model for the simulation of secondary electron emission,'' PRST-AB \textbf{5}, 124404 (2002).


\bibitem{ECLOUD12:FURMAN}
M.A.~Furman, ``Electron Cloud Effects in Accelerators,'' in Proc. of ECLOUD'12, La Biodola, Isola d'Elba, Italy,  June 5-8 2012, arXiv:1310.1706  [physics.acc-ph].


\bibitem{PhaseI}
``The C{\scriptsize ESR}TA: Phase I Report,'' Tech. Rep. CLNS-12-2084, LEPP, Cornell University, Ithaca, NY (Jan. 2013). http://www.lns.cornell.edu/public/CLNS/2012/

\bibitem{ICFA_News:ZIMMERMANN}
F. Zimmermann, \textit{et al.}, ICFA Beam Dynamics Newsletter No. 33, K. Ohmi \& M. Furman, Eds. (2004).


\bibitem{PAC05:MPPP031} T. Kroyer, F. Caspers, E. Mahner, ``The CERN SPS Experiment on Microwave Transmission Through the Beam Pipe,'' in Proc of PAC'05, MPPP031, Knoxville, TN, (2004).

\bibitem{PRL100:094801} S. De Santis, J. M. Byrd, F. Caspers, \textit{et al}, {\em Phys. Rev. Lett.} {\bf 100}, 094801 (Mar. 2008).

\bibitem{PRSTAB13:071002}  S. De Santis \textit{et al.},  {\em Phys. Rev. ST Accel. Beams} {\bf 13}, 071002 (Jul. 2010).

\bibitem{PRSTAB14:012802} S. Federmann, F. Caspers and E. Mahner,  {\em Phys. Rev. ST  \mbox{Accel.} Beams} {\bf 14}, 012802 (Jan. 2011).

\bibitem{ICFA_News:DUGAN}
G.F. Dugan, \textit{et al.}, ICFA Beam Dynamics Newsletter No.~50, J. Urakawa and W. Chou, Eds. (2009).

\bibitem{IPAC11:SIKORA_TEW}
J. P. Sikora, \textit{et al.}, `` Resonant TE Wave Measurements of Electron Cloud Densities at C{\scriptsize ESR}TA,'' in Proc. of IPAC'11, San {Sebasti\'{a}n}, Spain,  August 2011, TUPC170, p.1434, (2011).

\bibitem{ECLOUD12:SIKORA_TEW}
J. P. Sikora, \textit{et al.}, ``TE Wave Measurement and Modeling,'' in Proc. of ECLOUD'12, La Biodola, Isola d'Elba, Italy,  June 5-8 2012, arXiv:1307.4315  [physics.acc-ph].

\bibitem{IPAC13:Fermilab}
Y.-M. Shin, \textit{et al.}, ``Microwave Resonator Diagnostics of Electron Cloud Density Profile In High Intensity Proton Beam,'' in Proc. of IPAC'13, Shanghai, China, May 2013, MOPWA064, pp. 825-827,  (2013).

\bibitem{LAArzimovich1965:ElemPlasmaPhys} L. A. Arzimovich, {\em Elementary Plasma Physics}, Blaisdell Publishing Company, Waltham, Massachusetts (1965).

\bibitem{PR106:196} S. J. Buschbaum \& S. C. Brown, {\em Phys. Rev.} {\bf 106}, 196, (1957).

\bibitem{MAHeald1965:PlasDiagMicroW} M. A. Heald and C. B. Wharton, {\em Plasma Diagnostics with Microwaves}, John Wiley and Sons, New York (1965).

\bibitem{NIM14:SONNAD}
K.G. Sonnad, \textit{et al.}, Nuclear Instruments \& Methods in Physics Research A (2014), http://dx.doi.org/10.1016/j.nima.2014.03.052, Preprint, arXiv:1311.5946 [physics.acc-ph].

\bibitem{IPAC12:Alesini}
D. Alesini, \textit{et al.},``Experimental Measurements of e-Cloud Mitigation using Clearing Electrodes in the DAFNE Collider'', in Proc of IPAC'12, New Orleans, May 2012, TUOBC03, (2012).



\bibitem{KRSymon1971:Mechanics}
K.R. Symon, {\em Mechanics}, Addison-Wesley Publishing Company, Reading, Massachusetts (1971).

\bibitem{IPAC12:DESANTIS}
S. De Santis and J. P. Sikora,  \textit{et al}.,``Analysis of Modulation Signals Generated in the TE Wave Detection Method for Electron Cloud Measurements,'' in Proc. of IPAC'13, New Orleans, United States,  May 2012, MOPPR073, (2012).

\bibitem{IBIC13:DESANTIS}
S. De Santis and J. P. Sikora, ``Analysis of Modulation Signals Generated in the TE Wave Detection Method for Electron Cloud Measurements,'' in Proc. of IBIC'13, Oxford, United Kingdom, September 2013, TUPF36, (2013).

\bibitem{IBIC13:SIKORA}
J.P. Sikora, \textit{et al}., ``Resonant TE Wave Measurement of Electron Cloud Density Using Multiple Sidebands,'' in Proc.
of IBIC'13, Oxford, United Kingdom, September 2013, TUPF34, (2013).

\bibitem{IBIC13:phase_det}
J. P. Sikora, \textit{et al.}, ``Resonant TE Wave Measurements of Electron Cloud Density Using Phase Detection,'' in Proc. of IBIC'13, Oxford, United Kingdom, September 2013, TUPF35, (2013).

\bibitem{ELGinzton1957:MicroMeas} E. L. Ginzton, {\em Microwave Measurements}, McGraw-Hill Book Company, Inc, New York (1957).


\bibitem{IBIC12:L3_beadpull}
J. P. Sikora, \textit{et al.}, ``Electron Cloud Density Measurements Using Resonant TE Waves at C{\scriptsize ESR}TA,'' in Proc. of IBIC'12, Tsukuba, Japan, October 2012, TUPB49, (2012).


\bibitem{BeadPull:Slater}
L.C. Maier and J.C. Slater, J. Appl. Phys. {\textbf 23}, 68 (1952); doi: 10.1063/1.1701980


\bibitem{IEEE:Carter} R. G. Carter, ``Accuracy of Microwave Cavity Perturbation Measurements,'' {\em Microwave Theory and Techniques, IEEE Transactions on}, vol.{\bf 49}, no.5, pp.918-923, May 2001


\end{thebibliography}







\end{document}